\documentclass[
 reprint,
superscriptaddress,
 amsmath,amssymb,
 aps,
superscriptaddress,
 amsmath,amssymb,
 aps,longbibliography
]{revtex4-1}

\usepackage{graphicx}
\usepackage{dcolumn}
\usepackage{bm}
\usepackage{amsmath,amssymb,amstext,amsthm}
\usepackage{braket}
\usepackage{bbold}
\usepackage{bbm}
\usepackage{xcolor}
\usepackage{ gensymb }
\usepackage{soul}
\setstcolor{red}
\usepackage{ mathrsfs }
\usepackage{cancel}

\usepackage{hyperref}

\hypersetup{colorlinks,linkcolor=blue,citecolor=blue,urlcolor=blue}

\begin{document}

\preprint{APS/123-QED}
\setlength{\abovedisplayskip}{1pt}
\title{Generation of neutron Airy beams}

\author{Dusan Sarenac}
\email{dusansar@buffalo.edu}

\affiliation{Department of Physics, University at Buffalo, State University of New York, Buffalo, New York 14260, USA}

\author{Owen Lailey} 
\affiliation{Institute for Quantum Computing, University of Waterloo,  Waterloo, ON, Canada, N2L3G1}
\affiliation{Department of Physics and Astronomy, University of Waterloo, Waterloo, ON, Canada, N2L3G1}

\author{Melissa E. Henderson} 
\affiliation{Neutron Scattering Division, Oak Ridge National Laboratory, Oak Ridge, TN 37831, USA}

\author{Huseyin Ekinci} 
\affiliation{Institute for Quantum Computing, University of Waterloo,  Waterloo, ON, Canada, N2L3G1}
\affiliation{Department of Physics and Astronomy, University of Waterloo, Waterloo, ON, Canada, N2L3G1}
\author{Charles W. Clark}
\affiliation{Joint Quantum Institute, National Institute of Standards and Technology and University of Maryland, College Park, Maryland 20742, USA}
\author{David G. Cory}
\affiliation{Institute for Quantum Computing, University of Waterloo,  Waterloo, ON, Canada, N2L3G1}
\affiliation{Department of Chemistry, University of Waterloo, Waterloo, ON, Canada, N2L3G1}

\author{Lisa DeBeer-Schmitt} 
\affiliation{Neutron Scattering Division, Oak Ridge National Laboratory, Oak Ridge, TN 37831, USA}

\author{Michael G. Huber}
\affiliation{National Institute of Standards and Technology, Gaithersburg, Maryland 20899, USA}

\author{Jonathan S. White}
\affiliation{Laboratory for Neutron Scattering and Imaging, Paul Scherrer Institut, CH-5232 Villigen PSI, Switzerland}

\author{Kirill Zhernenkov}
\affiliation{J\"ulich Centre for Neutron Science at Heinz Maier-Leibnitz Zentrum, Forschungszentrum J\"ulich GmbH, 85748 Garching, Germany}

\author{Dmitry A. Pushin}
\email{dmitry.pushin@uwaterloo.ca}
\affiliation{Institute for Quantum Computing, University of Waterloo,  Waterloo, ON, Canada, N2L3G1}
\affiliation{Department of Physics and Astronomy, University of Waterloo, Waterloo, ON, Canada, N2L3G1}

\date{\today}


\pacs{Valid PACS appear here}


\begin{abstract}

The Airy wave packet is a solution to the potential-free Schr\"{o}dinger equation that exhibits remarkable properties such as self-acceleration, non-diffraction, and self-healing. Although Airy beams are now routinely realized with electromagnetic waves and electrons, the implementation with neutrons has remained elusive due to small transverse coherence lengths, low fluence rates, and the absence of neutron lenses. In this work, we overcome these challenges through a holographic approach and present the first experimental demonstration of neutron Airy beams. 
The presented techniques pave the way for fundamental physics studies with Airy beams of non-elementary particles, the development of novel neutron optics components, and the realization of neutron Airy-vortex beams. 


\end{abstract}
\maketitle

\section{Introduction}
In 1979, Berry and Balazs showed that there exists a solution to the potential-free Schr\"{o}dinger equation in the form of an Airy wave packet that is diffraction-free and manifests a form of self-acceleration~\cite{berry_Airy}.  Afterwards, 1D Airy states appeared in several hallmark neutron experiments, albeit always associated with a gravitational potential~\cite{greenberger_Airy,greenberger_overhauser1979,gibbs_quantum_bouncer,bound_neutron_propose, bound_neutron_exper, qubounce}. Greenberger offered a general physical interpretation of the nondispersive Airy wave packet according to Einstein's equivalence principle: the Airy wave packet is the stationary state solution for a free particle falling in a gravitational field, equivalent to a free falling reference frame in which the Schr\"{o}dinger equation is force-free~\cite{greenberger_Airy}. These concepts were applied to the Colella, Overhauser and Werner (COW) experiment, where a neutron interferometer was used for the first experimental observation of gravity's influence on a quantum particle~\cite{colella1975observation,greenberger_overhauser1979}. 
Airy functions also appeared in the `quantum bouncer' where a particle in a gravitational potential is reflected from a perfect mirror~\cite{gibbs_quantum_bouncer}, which led to the observation of bound neutron quantum states in Earth's gravitational potential~\cite{bound_neutron_propose, bound_neutron_exper, qubounce}. 

Over the last 20 years, significant progress has been made in the experimental generation and detection of free space propagating Airy beams and structured waves in general~\cite{rubinsztein2016roadmap, bliokh2023roadmap}. Optical Airy beams were first produced using a spatial light modulator that imprinted a cubic phase profile onto a coherent light beam and sent it through a lens that performed an optical Fourier transform to yield an Airy beam~\cite{airy_beam}. These optical systems confirmed the remarkable Airy wave properties of self-acceleration, non-diffraction, and self-healing, at specific propagation distances~\cite{airy_beam, airy_finite, airy_self_healing}. The results have proven useful in many applications~\cite{Airy_overview}, such as biomedical imaging to extend the depth of focus~\cite{Airy_microscopy, Airy_super_img, microscopy_Airy_SA}, generation of curved plasma channels~\cite{polynkin2009}, and for particle and current manipulation along curved trajectories~\cite{tweezers, ptcle_manip_Airy, Airy_nonparaxial, clerici2015}. Electron Airy beams have also been generated using a nanoscale cubic grating and a set of magnetic lenses~\cite{electron_airy}. 


\begin{figure*}
    \centering \includegraphics[width=1\linewidth]{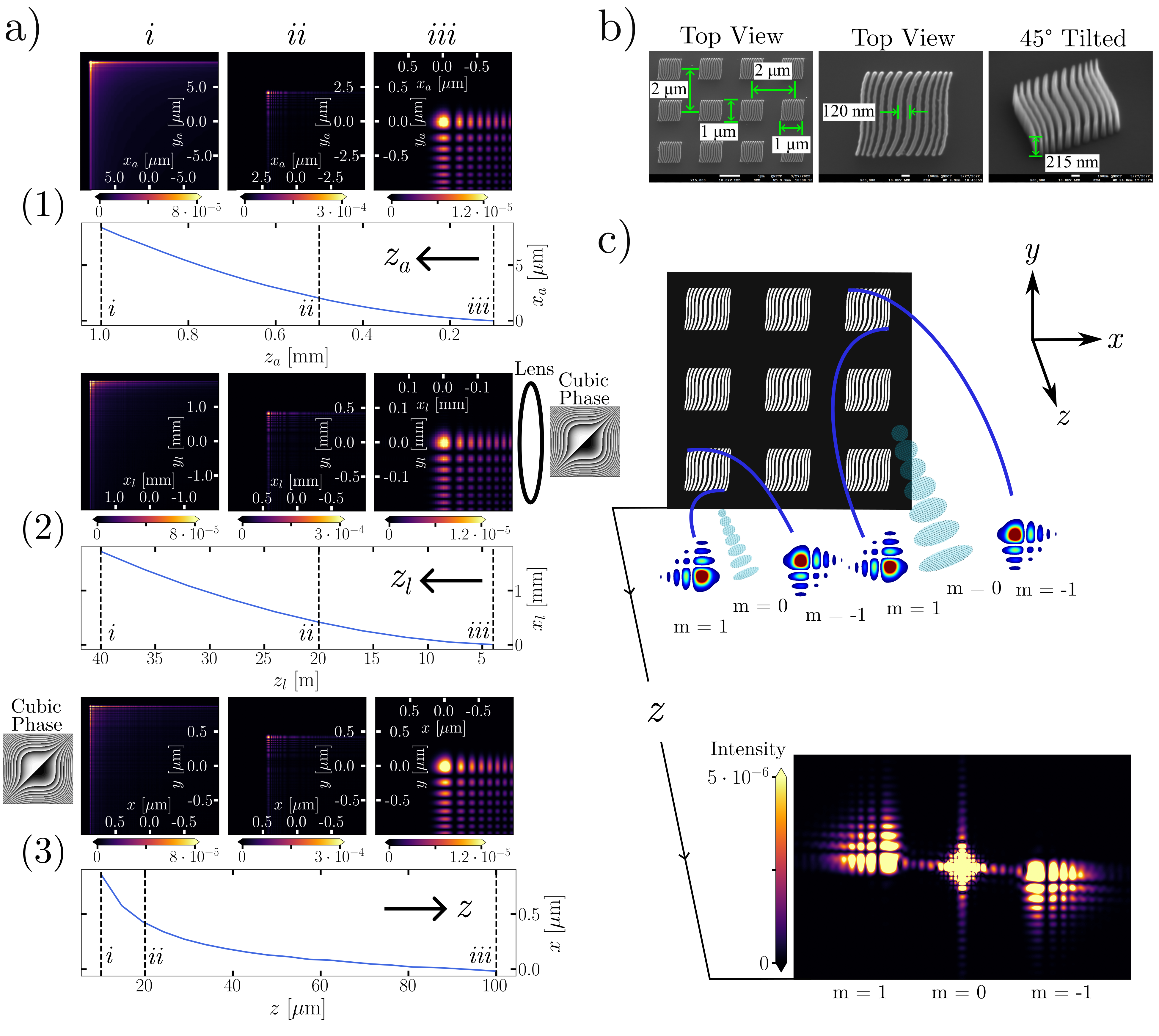}
    \caption{a) Three cases of Airy beam generation whereby the propagation dynamics can be equivalently mapped between each system. Case (1) considers the propagation of an inputted Airy wavefunction (Eq.~\ref{eq:input}) yielding intensity profiles (Eq.~\ref{eq:theory_intens}) at propagation distances of $z_{a,i} = 1.0$~mm, $z_{a, ii} = 0.5$~mm, and  $z_{a, iii} = 0.1$~mm, for $\lambda = 12$~\AA, $\sigma_{t} = 10~\mu$m and $x_0 \approx 0.1~\mu$m. Cases (2) and (3) consider practical Airy beam generation with a cubic phase mask with and without a lens respectively, for cubic coefficient $c = 50~\mu$m$^{-3}$, transverse coherence length $\sigma_{\perp} = 2~\mu$m, and focal length $f=20$~mm. The propagation distances are mapped to approximately: $z_{l,i} = 40$~m, $z_{l, ii} = 20$~m, $z_{l, iii} = 4$~m and $z_{i} = 10~\mu$m, $z_{ii} = 20~\mu$m, $z_{iii} = 100~\mu$m using Eq.~\ref{lens_intens} and Eq.~\ref{cubic_intens} respectively.
    b) SEM images characterizing the phase gratings used to experimentally generate neutron Airy beams. The array is $0.5$~cm by $0.5$~cm in size and consists of $6,250,000$ individual $1~\mu$m by $1~\mu$m phase gratings with period $120$~nm, height $300$~nm, and array period of $2~\mu$m. c) Pictorial depiction of neutron diffraction from the array of phase gratings which corresponds to case (3) of Airy beam generation. Incoming neutrons transmit through the phase gratings and produce a diffraction spectra with nonzero diffraction orders (m) exhibiting the well-defined Airy intensity patterns in the far field as seen in the simulation for $z = 12$~m, $\lambda = 12$~\AA, and $\sigma_{\perp} = 3~\mu$m. Near the grating, the trajectories of the nonzero diffraction orders cross over each other as indicated by the blue curves~\cite{electron_airy}.
    }
    \label{fig:concept}
\end{figure*}

In this work, we report the first experimental realization of free space propagating neutron Airy beams. This is achieved through a holographic approach that is compatible with existing neutron optics. 
We observe neutron Airy beam intensity profiles in the far field after transmission through a microfabricated array of cubic phase gratings and compare propagation dynamics to a free space neutron beam, using Small Angle Neutron Scattering (SANS) techniques. This work extends the toolbox of Airy beams to non-elementary particles and opens up possibilities for fundamental physics experiments as well as novel techniques for shaping of the neutron wavefunction.

\newpage
\section{Airy Beam Dynamics}
\label{section2}
The Airy function $\operatorname{Ai}(s)$ can be defined as \cite{Olver:2010:NHMF}:

\begin{equation}
\operatorname{Ai}\left(s\right)=\frac{1}{\pi}\int_{0}^{\infty}\cos\left(\tfrac%
{1}{3}t^{3}+st\right)\,\mathrm{d}t,
\label{airy_func}
\end{equation}

\noindent where $s=x_a / x_0$ is a dimensionless Cartesian coordinate for rectilinear wave propagation, $x_0$ is the characteristic length scale, and $t$ is an integration variable. The maximum probability Airy lobe trajectory in the transverse plane is given by: 

\begin{equation}
    x_a(z_a) = \frac{1}{4k^2x_0^3} z_a^2,
    \label{eq:traj}
\end{equation}

\noindent where $z_a$ is propagation distance and $k$ is the beam's wave vector~\cite{airy_finite, airy_beam}. The pure Airy function is nonphysical as it has infinite energy, and thus in practice is truncated with an aperture function such as an exponential aperture~\cite{airy_finite} or with a Gaussian envelope~\cite{Sanz2024}. The wavefunction at $z_a=0$ for a Gaussian truncated 2D Airy beam is thus given by:

\begin{equation}
    \psi_i(x_a, y_a, 0) = \mathcal{N} e^{-\frac{x_a^2 + y_a^2}{2\sigma_{t}^2}} Ai \left(\frac{x_a}{x_0} \right) Ai \left(\frac{y_a}{y_0} \right),
    \label{eq:input}
\end{equation}

\noindent where $\sigma_{t}$ is the envelope's standard deviation, $\mathcal{N}$ is a normalization constant, and setting $y_0 = x_0$ gives identical propagation dynamics in $x_a$ and $y_a$. The evolution of the wavefunction as it propagates a distance $z_a$ to the detector can be computed with the Fresnel-Kirchoff integral yielding $\psi_f(x_a', y_a', z_a)$, where $x_a',\: y_a'$ are transverse coordinates at the detector. The intensity distribution is described by:

\begin{equation}
    I_a(x_a', y_a', z_a) = |\psi_f(x_a', y_a', z_a)|^2.
    \label{eq:theory_intens}
\end{equation}

The intensity profiles shown in case (1) of Fig.~\ref{fig:concept}a correspond to propagating the wavefunction $\psi_i(x_a, y_a, 0)$ a distance of $z_{a, i} = 1.0$~mm, $z_{a, ii} = 0.5$~mm, and $z_{a, iii} = 0.1$~mm, for $\lambda = 12$~\AA, $\sigma_{t} = 10~\mu$m and $x_0 \approx 0.1~\mu$m. At $z_{a, ii} = 0.5$~mm, the Airy beam has shifted in the transverse plane according to Eq.~\ref{eq:traj} and at $z_{a, i} = 1.0$~mm, the characteristic structure of the Airy beam is just beginning to degrade due to the Gaussian truncation. 

In practice, Airy beams can be generated experimentally by imprinting a cubic phase $\phi(x, y) = c_xx^3 + c_yy^3$, where typically $c_x = c_y\equiv c$, on a Gaussian wave packet described by transverse coherence length $\sigma_{\perp}$ and then performing an optical Fourier transform by using a lens~\cite{airy_finite, airy_beam}. The propagation dynamics of the Airy beam after the focal spot are scaled relative to the case (1) described above, depending on focal length $f$ and the implicit relationship between $c$ and $x_0$:

\begin{equation}
    I_l(x_l, y_l, z_l) = I_a \left( \frac{a}{\lambda f}x_l, \frac{a}{\lambda f}y_l, \left( \frac{a}{\lambda f} \right)^2 z_l \right),
    \label{lens_intens}
\end{equation}

\noindent where $I_l(x_l, y_l, z_l)$ is the intensity after the focal spot as shown in Fig.~\ref{fig:concept}a case (2), $I_a$ is given by Eq.~\ref{eq:theory_intens}, and $a = \sqrt{6\pi}x_0 / c^{1/3}$. It follows that Eq.~\ref{eq:traj} also gets scaled:

\begin{equation}
    x_l(z_l) = \frac{a^3}{4k^2x_0^3\lambda^3}\frac{z_l^2}{f^3}.
    \label{cubic_lens_traj}
\end{equation}
With $\sigma_{\perp} = 2~\mu$m, $f = 20~$mm, and $c=50~\mu$m in Fig.~\ref{fig:concept}a case (2), we obtain parabolic deviation of a few mm over tens of meters, corresponding to an acceleration of about $0.24$~m/s$^2$. 

Here we require an alternative approach for generating Airy beams in order to circumvent the challenges associated with neutron beams. Our approach is to imprint a cubic phase on a Gaussian wave packet and let the beam freely propagate without employing a lens. According to the Fraunhofer approximation, the wavefunction in the far field is well approximated by the Fourier transform of the cubic phase profile. We therefore achieve the Airy wavefunction in the far field with an inverse relationship in the propagation dynamics compared to the Airy beam generation with a lens, as shown in Fig.~\ref{fig:concept}a case (3). The intensity profile after the cubic phase mask is given by:

\begin{equation}
    I(x, y, z) = I_a \left( \frac{a}{\lambda z}x, \frac{a}{\lambda z}y, \left( \frac{a}{\lambda z} \right)^2 z \right).
    \label{cubic_intens}
\end{equation}

\noindent Applying this transformation to Eq.~\ref{eq:traj}, we see that the typical parabolic trajectory associated with Airy beam propagation is mapped to an inverse $z$ relationship:

\begin{equation}
    x(z) = \frac{a^3}{4k^2x_0^3\lambda^3}\frac{1}{z}.
    \label{cubic_traj}
\end{equation}

This approach is well-suited for neutrons as it circumvents the fact that the neutron index of refraction in common materials differs from the vacuum by order only $10^{-5}$, making neutron lens fabrication currently impractical. 

\begin{figure*}[ht]
    \centering\includegraphics[width=1\linewidth]{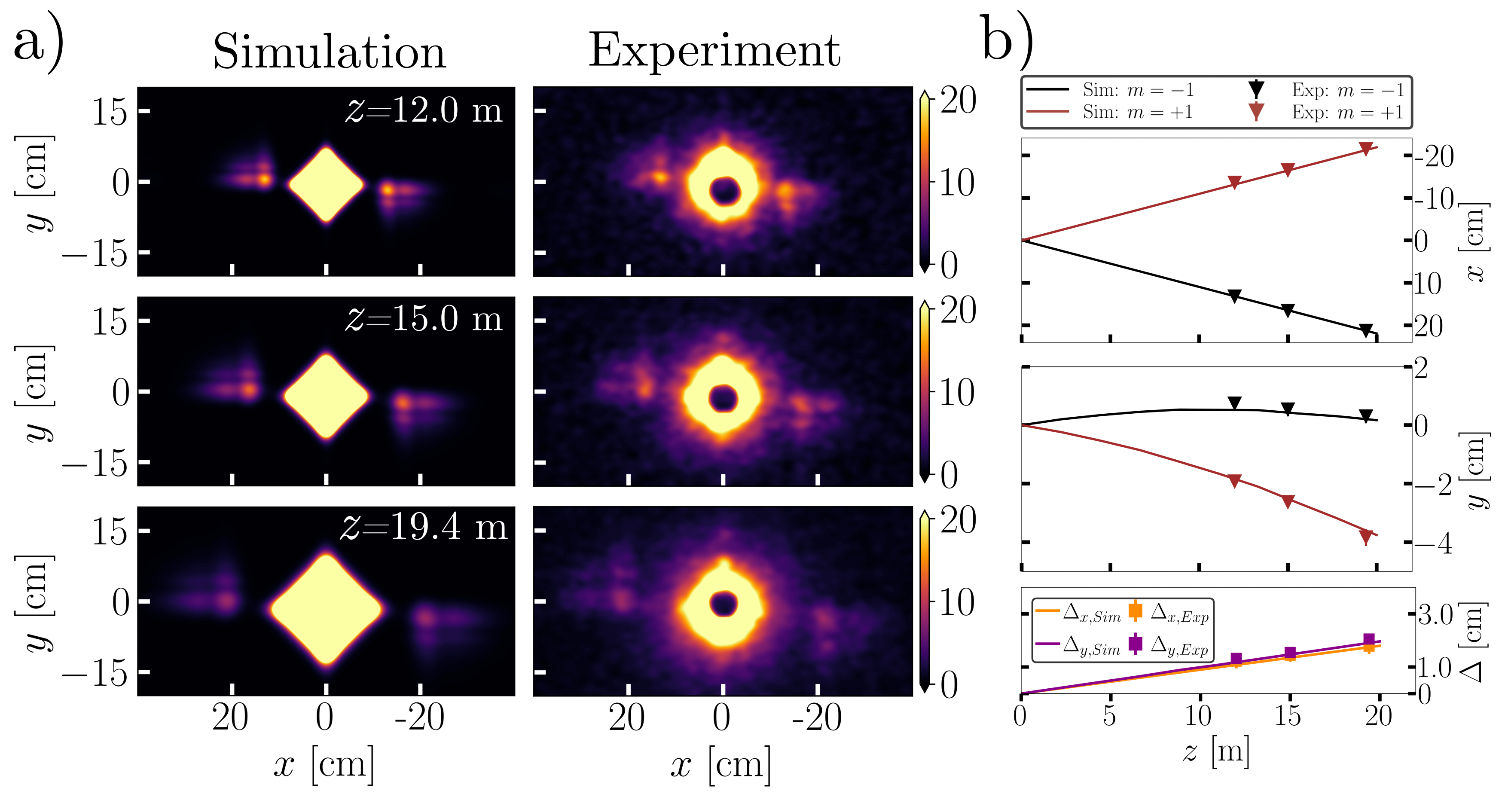}
    \caption{a) The measured Airy diffraction spectra observed at detector positions of $z = 12.0$~m, $15.0$~m, and $19.4$~m. A beam trap was placed at the center of the detector in the direct beam path to better emphasize neutron counts of the nonzero diffraction orders at the detector. We find good agreement with simulation that considers a transverse coherence length of $\sigma_{\perp} = 3~\mu$m and we include all of the experimental parameters in the simulation, other than the beam trap, such as wavelength distribution, phase grating array structure, and gravity. b) Propagation dynamics of the Airy first diffraction orders in both $x$ and $y$ directions. To obtain the trajectory of the main Airy lobe, and error bars, we fit integrated experimental data to 1D Airy functions given by Eq.~\ref{airy_func} in $x$ and $y$ directions and find the maximum intensity coordinate. To compare Airy and linear propagation dynamics, we determine the $x$- and $y$-displacement $(\Delta_x,\Delta_y)$ of the maximum intensity position of the Airy beam relative to the position of the linear diffraction spots and observe linear separation between the curves. Note that in b) some error bars lie within the experimental data markers. 
    }
    \label{fig:data}
\end{figure*}

\section{Materials and Methods}

An array of binary diffraction phase gratings was fabricated on a silicon wafer where each individual grating profile is given by:

\begin{equation}
    \frac{1}{2} \left( \text{sgn} \left[\cos \left\{ \frac{2\pi}{p}x + c_xx^3 - c_yy^3  \right\} \right] + 1 \right),
    \label{eq:phase_prof}
\end{equation}

\noindent where $2\pi/p$ is the carrier frequency and $c_x,~c_y$ are cubic coefficients of the grating. The array covered a 0.5 cm by 0.5 cm area and consisted of $6,250,000$ individual 1 $\mu$m by 1 $\mu$m phase gratings, where each one possessed a period $p = 120$~nm, height $300$~nm, $c_x = 55~\mu$m$^{-3}$, $c_y = 45~\mu$m$^{-3}$, and was separated by 1~$\mu$m on each side from the other phase gratings, as shown in Fig.~\ref{fig:concept}b. The fabrication procedure is identical to the fork-dislocation phase grating procedure outlined in the Supplementary Material of Ref.~\cite{sarenac2022experimental}. 


A preliminary study was done at the SANS-I beamline at the Paul Scherrer Institute~\cite{kohlbrecher2000new}, as a result of which an improved setup was devised and implemented at the GP-SANS beamline at the High Flux Isotope Reactor at Oak Ridge National Laboratory~\cite{wignall201240}. The silicon wafer was placed inside a mount 17.8 m away from a 20 mm diameter source aperture. Directly in front of the sample was a 4~mm diameter sample aperture. We varied the distance from the sample to the detector from $12.0$~m to $19.4$~m. At each distance, we took measurements with the cubic phase grating array (schematic in Fig.~\ref{fig:concept}c), as well as a linear phase grating array where $c_x = c_y = 0$ in Eq.~\ref{eq:phase_prof} (see the Supplementary Material of Ref.~\cite{sarenac2022experimental} for the SEM profiles of the linear phase gratings). The detector spans an area of $\approx 1$ m$^2$ with each pixel being $\approx 5.5$ mm by $4.3$ mm in size. The wavelength distribution was triangular with $\Delta\lambda/\lambda\approx0.13$, where $\Delta\lambda$ is the FWHM and the central wavelength is $12$~\AA. The final SANS data was passed though a low-pass filter to remove the Poissonian noise that varies from pixel to pixel.

\section{Results}
\label{results}

The experiment consisted of measuring far field intensity from an array of cubic and linear phase gratings at grating to camera distances of: $z=12.0$~m, $15.0$~m and $19.4$~m. Fig.~\ref{fig:data}a shows the observed Airy beams at different $z$ distances (see Appendix for linear phase grating reference measurements). For the simulated profiles, we apply the cubic phase array, with each individual grating profile given by Eq.~\ref{eq:phase_prof}, to a neutron wave packet with transverse coherence length of $\sigma_{\perp} = 3~\mu$m at the grating and propagate to the far field, as discussed in detail in section~\ref{section2}. We include all of the experimental parameters in the simulation such as wavelength distribution, phase grating array structure, gravity, etc., and in addition account for the experimental resolution by applying a low-pass filter. To account for experimental effects that cause smearing of the intensity pattern, such as vibrations, temperature fluctuations, array size, and pixel size, we convolve the simulated images at each detector position $z$ with a Gaussian described by $\sigma_{c}(z) = 1.1$~cm $(z/19.4$~m$)$, where $\sigma_{c}(19.4$~m$) = 1.1$~cm is obtained by minimizing the difference between the experimental linear diffraction spot size and the simulation at $z = 19.4$~m. 

In Fig.~\ref{fig:data}b we compute the maximum intensity coordinates of the main Airy lobe and linear diffraction spots for $m = \pm 1$ to observe the propagation dynamics and find good agreement with simulations. In the $y$ direction, Airy maximum intensity coordinates follow a parabolic trajectory due to the gravitational force; in the $x$ direction, they linearly separate in this far field regime.

To further examine Airy propagation dynamics, we can determine the $x$- and $y$-displacement $(\Delta_x,\Delta_y)$ of the maximum intensity position of the Airy beam $(x_A,y_A)$ relative to the propagation center of the first diffraction order $(x_L,y_L)$:

\begin{equation}
    \Delta_{x,y} = \frac{1}{2}\left|A_{x,y} - L_{x,y}\right|,
\end{equation}

\noindent whereby the displacement is calculated by considering the separation distances $(A_{x,y},L_{x,y})$ between the $m=1$ and $m=-1$ diffraction orders for both the Airy and linear case respectively:



\begin{equation}
    A_{x,y} = \left|\{x_A,y_A\}_{m = +1} - \{x_A,y_A\}_{m = -1}\right|,
\end{equation}

\begin{equation}
    L_{x,y} = \left|\{x_L,y_L\}_{m = +1} - \{x_L,y_L\}_{m = -1}\right|.
\end{equation}

\noindent As shown in Fig.~\ref{fig:data}b, we observe linear separation between the curves in both $x$ (orange curve) and $y$ (purple curve) transverse coordinates, which is in good agreement with theory. 

\section{Conclusion and Discussion}

We have introduced and experimentally demonstrated the generation and detection of neutron Airy beams. A holographic approach was used to imprint a cubic phase profile on the neutron wavefunction and observe the Airy beam formation in the far field. We compared Airy beam propagation dynamics with linear diffraction in this far field regime and found excellent agreement with simulations. This work paves the way for further exploration of fundamental properties of Airy beams with neutrons; for example, neutrons are well-suited to study the self-healing properties of Airy beams in scattering experiments, as was done with light in a sample of mono-disperse silica microspheres~\cite{airy_self_healing}, since samples of dilute hard spheres are commonly used in neutron imaging and grating interferometer experiments~\cite{andersson2008analysis, spheres_dfi, sarenac2024cone}.


Several exciting applications stem from this work; for example, there is recent interest in Airy-vortex beams, whereby a helical wave carrying Orbital Angular Momentum (OAM) is superimposed with an Airy beam~\cite{airy_vortex, airy_vortex1, airy_vortex2, karlovets2015}, and the unique interaction of these beams with chiral media~\cite{airy_chiral, airy_vortex_chiral}. Using techniques from Refs.~\cite{sarenac2022experimental, oam_phase}, neutron Airy-vortex beams could be used to study the scattering properties of Skyrmion samples~\cite{henderson2021characterization, henderson2022skyrmion,henderson2023three}. Moreover, the coherent superposition of counterpropagating Airy beams has been shown to be abruptly autofocusing along the propagation axis, increasing intensity by orders of magnitude, and then exhibiting Young-type interference fringes with further propagation~\cite{auto_focus_theory, auto_focus_exp, airy_young_fringes}. This is attractive for neutron experiments where the common optical element, a lens, is impractical and thus abruptly autofocusing neutron Airy beams could be used to improve contrast in neutron imaging, for example. Lastly, this work could also be useful for the experimental investigation of accelerating wave packets accumulating a geometric phase (Berry- or Aharonov-Bohm-like) in a system with no potential whatsoever~\cite{geometric_phase_airy, AB_neutrons}. 

\section*{Acknowledgements}

This work was supported by the Canadian Excellence Research Chairs (CERC) program, the Natural Sciences and Engineering Research Council of Canada (NSERC), the Canada  First  Research  Excellence  Fund  (CFREF), and the US Department of Energy, Office of Nuclear Physics, under Interagency Agreement 89243019SSC000025. This work was also supported by the DOE Office of Science, Office of Basic Energy Sciences, in the program "Quantum Horizons: QIS Research and Innovation for Nuclear Science" through grant DE-SC0023695. A portion of this research used resources at the High Flux Isotope Reactor, a DOE Office of Science User Facility operated by the Oak Ridge National Laboratory. This work is based partly on experiments performed at the Swiss spallation neutron source SINQ, Paul Scherrer Institute, Villigen, Switzerland.




\bibliography{mybib}

\clearpage
\onecolumngrid

\section*{Appendix}
\begin{figure*}[ht]
    \centering\includegraphics[width=0.75\linewidth]{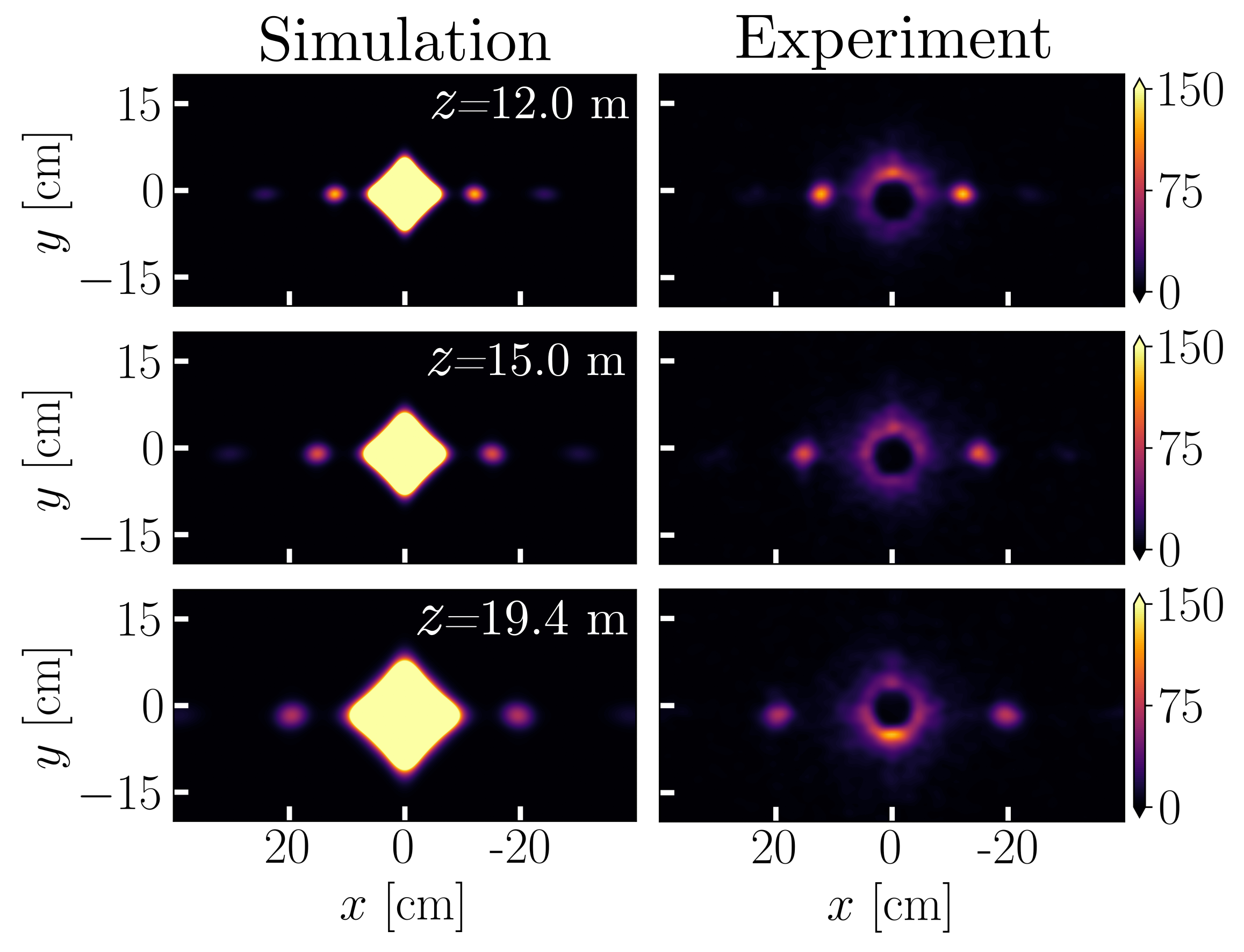}
    \caption{The measured linear diffraction spectra observed at detector positions of $z = 12.0$~m, $15.0$~m, and $19.4$~m. A beam trap is placed at the center of the detector in the direct beam path to better emphasize neutron counts of the nonzero diffraction orders at the detector. We find good agreement with simulation that considers a transverse coherence length of $\sigma_{\perp} = 3~\mu$m and we include all of the experimental parameters in the simulation, other than the beam trap, such as wavelength distribution, phase grating array structure, and gravity.
    }
    \label{fig:linear}
\end{figure*}

The measurement of far field neutron intensity after an array of linear phase gratings is used as a reference point to determine neutron Airy beam propagation dynamics. Shown in Fig.~\ref{fig:linear} are the measured linear phase grating benchmark measurements at grating to camera distances of: $z=12.0$~m, $15.0$~m and $19.4$~m. We find good agreement with simulation and compare the linear and Airy beam propagation dynamics in section~\ref{results} and Fig.~\ref{fig:data}b.

\end{document}